\documentclass[prl,twocolumn,showpacs,amsmath,amssymb,superscriptaddress,floatfix]{revtex4-1}

\usepackage{calc}
\usepackage{float}
\usepackage{graphicx}
\usepackage{bm}
\usepackage[percent]{overpic} 
\usepackage[colorlinks=true]{hyperref} 
\usepackage{amsmath}
\usepackage{amssymb}
\usepackage{latexsym}
\usepackage{multirow}
\usepackage{soul}

\newcommand{\abs}[1]{\left|\,#1\,\right|} 
\newcommand{\avg}[1]{\left< #1 \right>} 

\newcommand{\kommentar}[1]{}

\begin{document}
\title{Complex Quantum Networks: From Universal Breakdown to Optimal Transport}
\author{Oliver M\"ulken}
\email{muelken@physik.uni-freiburg.de}
\author{Maxim Dolgushev}
\affiliation{Physikalisches Institut, Universit\"at Freiburg, Hermann-Herder-Str.\ 3, D-79104 Freiburg, Germany}
\author{Mircea Galiceanu}
\affiliation{Departamento de F\'isica, Universidade Federal do Amazonas,
3000, Japiim, 69077-000, Manaus-AM, Brazil}
\affiliation{Institut f\"ur Theoretische Physik, Technische Universt\"at Dresden, 01062 Dresden, Germany}

\date{\today} 

\begin{abstract} 
We study the transport efficiency of excitations on complex quantum networks with loops. 
For this we consider sequentially growing networks with different topologies of the sequential subgraphs.
This can lead either to a universal complete breakdown of transport for complete-graph-like sequential subgraphs or to optimal transport for ring-like sequential subgraphs. 
The transition to optimal transport can be triggered by systematically reducing the number of loops of complete-graph-like sequential subgraphs in a small-world procedure.
These effects are explained on the basis of the spectral properties of the network's Hamiltonian. Our theoretical considerations are supported by numerical Monte-Carlo simulations for complex quantum networks with a  scale-free size distribution of sequential subgraphs and a small-world-type transition to optimal transport.
\end{abstract}


\maketitle
\date{\currenttime}

{\it Introduction.---}
Complex networks are a beautiful tool to understand the statistics and dynamics of a huge variety of systems, such as biological systems, social groups, or the Internet, see, for instance, \cite{newman2006structure}. Many of these networks can be grouped into classes, for instance, scale-free networks \cite{barabasi2009scale} or small-world networks \cite{newman1999scaling}, where these intrinsic properties of the link (bond) distribution between the nodes determine the  {\it classical} static \cite{dorogovtsev2013evolution} as well the {\it classical} dynamic features \cite{barzel2013universality}. 
Only recently, network theory has been combined with quantum theory, in order to study, say, the quantum dynamic properties on complex structures \cite{mulken2011continuous, sanchez2012quantum, garnerone2012adiabatic, bianconi2015interdisciplinary, bianconi2015complex, faccin2013degree, faccin2014community}.

The majority of networks will have (some) loops, which ---for classical networks--- influence the dynamics. For instance, the target search on looped DNA is of super-diffusive type \cite{lomholt2005optimal}. In the cell, DNA appears as supercoils (plectonemes), which also influences the dynamics \cite{van2012dynamics}. It is not clear, if and how the presence of loops influences the  {\it quantum} dynamics. For the subclass of quantum networks without loops, we have recently demonstrated that there are universal features when the complexity of the network leads to a complete breakdown of the quantum transport properties \cite{kulvelis2015universality}.

Complex quantum networks appear in quantum information theory, e.g., in the study of quantum decision trees \cite{farhi1998quantum} or of quantum search engine ranking \cite{garnerone2012adiabatic}, where loops can be present. In nature one might encounter such networks, e.g., in the assemblies of ring-like LH1 and LH2 complexes in bacterial light-harvesting antennae, where quantum dynamical aspects (can) play an important role for the efficiency of the process even at room temperature \cite{scholes2011lessons,hildner2013quantum}. Recently, coherent energy transfer at room temperature has also been shown for optimized architectures of artificial supramolecular nanofibers \cite{haedler2015long}. 

In this Letter, we focus on the influence of loops on the dynamics on sequentially growing complex quantum networks. 
By manipulating the structure of the sequential subgraphs (SSGs), we are able to induce a transition to optimal transport, characterized by a global time-averaged efficiency measure.

{\it Quantum Transport on Networks.---}
A network (undirected graph) $G=G(N,M)$ is defined by its $N$ nodes (vertices) and $M$ bonds (edges) \cite{brouwer2011spectra}. 
Each node is represented by a state $|j\rangle$, $j=1,\dots,N$. The quantum dynamics on such networks is governed by its 
Hamiltonian $\bm{H}$, which reflects the topological structure of the network, i.e., whenever two nodes $k$ and $j$ are connected by a single bond, one has in the node representation that $H_{k,j} \equiv \langle k | \bm{H} | j\rangle = \text{const}$, which can be chosen to be one without loss of generality. There is some freedom in choosing the diagonal elements $H_{j,j}$, we assume that these elements are a function of the degree $f_j$ of node $j$, i.e., $H_{j,j} = H(f_j)$. This includes the adjacency matrix $\bm{A}$, $H(f_j)=0$ for all $j$, as well as the connectivity matrix (Laplacian) $\bm{C}$, $H(f_j) = f_j$ \cite{mulken2011continuous}. The time-dependent transition amplitudes $\alpha_{kj}(t) = \langle k| \exp(-i \bm{H} t) | j \rangle$ and the corresponding transition probabilities $\pi_{kj}(t) = \abs{\alpha_{kj}(t)}^2$.

In order to quantify a network's transport efficiency, we use the space averaged probability $\overline\pi(t) \equiv \sum_j \pi_{j,j}(t)/N$ as a measure. If this quantity is small (large) for almost all times, the probability to leave any node of the network is ---on average--- large (small) rendering transport (in)efficient. By taking the long-time average of $\overline\pi(t)$, we arrive at a global time-independet measure for the transport efficiency \cite{mulken2011continuous}:
\begin{equation}
\overline\chi \equiv \lim_{t\to\infty}\frac{1}{t}\int_{0}^{t} dt'\overline{\pi}(t').
\end{equation}
In the spectral decomposition, $\overline\pi(t)$ as well as $\overline\chi$ depend on the eigenstates of $\bm{H}$, such that we employ the Cauchy-Schwarz inequality to obtain measures which solely depend on the spectral density $\varrho(E)$ of $\bm{H}$: one has $\overline\pi(t) \geq \abs{\overline\alpha(t)}^2$ with $\overline\alpha(t) = \sum_j \alpha_{j,j}(t)/N$ and \cite{kulvelis2015universality}
\begin{eqnarray}
\overline\chi \geq \chi &\equiv& \lim_{t\to\infty}\frac{1}{t}\int_{0}^{t}dt'\abs{\bar{\alpha}(t')}^2 = \sum_E \varrho^2(E)  \\
&\geq& \varrho^2(E_*) + \frac{1}{N}[1-\varrho(E_*)] \equiv \underline\chi,
\label{eq.chi}
\end{eqnarray}
where $E_*$ is a, at this point arbitrarily chosen, single eigenvalue.
Now we are in the position of discussing the quantum transport efficiency on networks on the basis of the spectral density of $\bm{H}$.
However, the number of highly degenerated eigenvalues depends on the choice of $\bm{H}$. A single highly degenerate eigenvalue $E_*$ can be present for one of the SSGs (few, when there are SSGs of the same size) when $\bm{H}=\bm{C}$ or for many of the SSGs when $\bm{H}=\bm{A}$.

\begin{figure}[b]
\centerline{\includegraphics[width=\columnwidth]{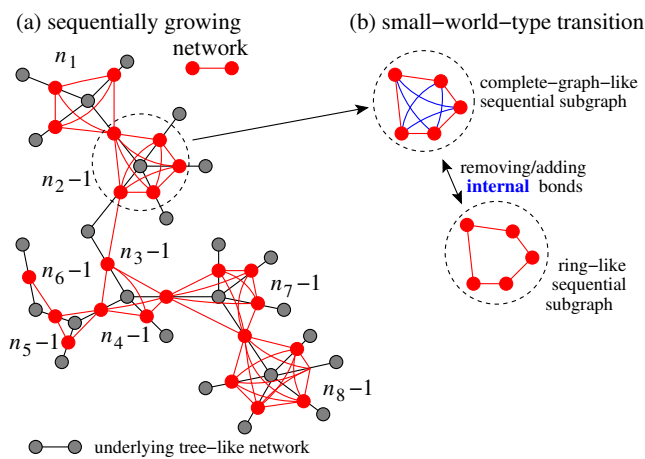}}
	\caption{(color online) (a) Sketch of a sequentially growing network with $g=8$ SSGs and the tree-like backbone. (b) Small-World-type transition in a SSG, where internal bonds (blue) are randomly added (removed) from the ring-like (complete-graph-like) SSGs with a certain probability $p_{\text{int}}$. 
}
\label{fig.tree-dual}
\end{figure}

{\it Sequentially Growing Networks.---}
In the following, we will consider sequentially growing networks of total size $N$, defined by a probability distribution, $p(n)$ (with $n\geq2$), for the size $n$ of a SSG \cite{galiceanu2012relaxation}. Specifically, we draw a number $n_1$ according to $p(n)$. This determines the size of the first SSG with a number of loops of the order $n_1$, but independent of the actual topology of this SSG. 
We then randomly pick one of the $n_1$ nodes and attach to this another SSG (with a similar topology) of size $n_2-1$. From the remaining $n_1+n_2-2$ nodes we again pick randomly a node to which we attach the next SSG of size $n_3-1$. In this way only at most two SSGs are connected at a given node. This procedure continues until we reach $N=\sum_{k=1}^g n_k - (g-1)$, where $g$ is the number of SSGs in the sequence, see also Fig.~\ref{fig.tree-dual}(a).
For finite $N$, the average number of SSGs, $\avg{g}$, 
can be obtained from $p(n)$ and the average size $\avg{n}$ of a SSG by
$\avg{g} = (N-1)/(\avg{n}-1)$.

The probability distribution can show a parameter dependence, e.g., for scale-free distributions (see below), such that one can interpolate between (i) small number of large SSGs, i.e., large $\avg{n}$, and (ii) larger numbers of small SSGs, i.e., small $\avg{n}$. In case (i) the topology of the SSGs can lead to highly degenerate eigenvalues, in particular, it is possible to find a single highly degenerate eigenvalue $E_*$. For case (ii), one can obtain a flat density $\varrho(E)$. Taking the average size $\avg{n}$ as a parameter dependent measure, we assume that $\varrho(E_*)=\varrho(E_*,\avg{n})$, such that $\varrho(E_*,\avg{n})\to1$ for $\avg{n}\to N$ and $\varrho(E_*,\avg{n})\to1/N$ for $\avg{n}\to2$ (two is the lower bound for connected networks).
This will also render $\underline\chi=\underline\chi(\avg{n})$, from which we will define the order parameter $1-\underline\chi(\avg{n})$ \cite{kulvelis2015universality}. In the $N\to\infty$ limit, this order parameter, denoted by $1-\underline\chi_\infty$, is one for optimal transport (when $\avg{n}\to2$), e.g., for rings or chains of infinite size, and zero for complete absence of transport (when $\avg{n}\to\infty$), e.g., for networks with a single eigenvalue with spectral density of order one such as star graphs or complete graphs. 
The breakdown of transport for $\avg{n}\to\infty$ is captured by a critical exponent
\begin{equation}
\underline\kappa \equiv \lim_{\avg{n}\to\infty} {\log [1-\underline\chi_\infty(\avg{n})]}\Big/{\log [1/{\avg{n}}]},
\end{equation}
which is bounded from above. 
For random trees, which are also sequentially growing networks, but without loops, the spectral density is related to the average degree $\avg{f}$ of those nodes with degree larger than one. For a node in a tree with degree $f>1$, the size of the corresponding SSG is $n=f+1$, which counts all nodes connected by a single bond to that particular node with $f>1$. 
In the limit of diverging average degree, 
one finds $\varrho(E_*,\sigma\simeq\sigma_c) \approx 1 - 2/(\avg{f}-1)$ \cite{kulvelis2015universality}, leading to $\underline\kappa=1$. Since all other networks will have smaller densities of the most highly degenerate eigenvalue, $\underline\kappa$ is bounded from above by one. 
In the case of sequentially growing networks with loops and a single highly degenerate eigenvalue $E_*$, one has $1-\avg{\varrho(E_*,\avg{n}\simeq N)} \sim 1/\avg{n}$. Then, $\underline\kappa$ is also bounded from below also by one, since 
\begin{equation}
1- \avg{\underline\chi} = 1- \avg{\varrho^2(E_*,\avg{n})} \leq 1- \avg{\varrho(E_*,\avg{n})}^2 \sim 1/\avg{n}.
\end{equation}
Replacing $\avg{n}$ by the size of that SSG with the maximally degenerate eigenvalue, $n_{\text{max}}$, for $\bm{H}=\bm{C}$, yields a similar result.

{\it Transition to Optimal Transport.---}
Clearly, transport not always breaks down completely if the size of a SSG becomes of the order of $N$. An additional requirement is that this SSG leads to a single highly degenerate eigenvalue. If two or more eigenvalues are highly degenerate, it directly follows from Eq.~(\ref{eq.chi}) that $\chi<1$ since two or more spectral densities in the sum are strictly smaller than one \cite{mulken2011continuous}. 

For complex networks with loops, we start from complete-graph-like SSGs. By replacing every complete-graph-like SSG of size $n>3$ by a ring of the same size (for $n=3$ the complete graph is a triangle which is also a ring), we obtain a network which is, depending on the probability distribution, a collection of connected ring-like SSGs. 
It is believed that regularities play an important role, see, e.g., Ch.\ 15 of Ref.~\cite{brouwer2011spectra}, and that they are the reason for large degeneracies of eigenvalues, however, no rigorous proof seems to exist. For our initially built complex networks, all nodes in a complete-graph-like SSG have the same distance from that node, which connects to other parts of the network, and thus are considered to be identical (indistinguishable). For networks with ring-like SSGs this is not the case, because one can distinguish the nodes on the ring-like SSG by distance from the node connecting to the rest of the network.

Sequentially growing networks have an underlying tree-like backbone if one only considers the connections between SSGs. If the SSGs are complete-graph-like, the complex network with $N$ nodes can be viewed as the dual network, $d(G)$, of a tree, $G$, with $N+1$ nodes, see also Fig.~\ref{fig.tree-dual}(a).
Then, one can obtain the spectrum of $\bm{H}[d(G)] = \bm{A}[d(G)]$ directly from the spectrum of the connectivity matrix $\bm{C}(G)$ of the tree, using the incidence matrix $\bm{B}(G)$, which relates nodes and bonds \cite{brouwer2011spectra}:
$\bm{C}(G) = \bm{B}(G) \bm{B}^T(G)$,
where $\bm{B}^T(G)$ is the transposed matrix. 
We note that $\bm{B}(G)$ is not symmetric, for trees with $N+1$ nodes there are $N$ bonds, such that it is a $(N+1)\times N$ matrix.
Also, the adjacency matrix $\bm{A}[d(G)]$ of the dual network follows from $\bm{B}(G)$ as
$\bm{A}[d(G)] = \bm{B}^T(G) \bm{B}(G) - 2 \bm{I}$,
where $\bm{I}$ is the identity matrix. It follows that the spectrum 
${\rm spec}[\bm{C}(G)] \sim {\rm spec}[\bm{A}[d(G)]+2 \bm{I}] \backslash \{0\}$ \cite{brouwer2011spectra}. 
The term $2\bm{I}$ only accounts for a constant shift in the spectrum of $\bm{A}[d(G)]$. Therefore, the spectral densities of $\bm{C}(G)$ and $\bm{A}[d(G)]$ are equivalent. If $\bm{C}(G)$ leads to highly degenerate eigenvalues, so will $\bm{A}[d(G)]$.
We note that choosing $\bm{H}[d(G)] = \bm{C}[d(G)]$, results also in highly degenerate eigenvalues but each complete-graph-like SSG has its ``own'' highly degenerate eigenvalue since a single complete graph of size $n$ has a $(n-1)$-fold degenerate eigenvalue $E_*=n$. Thus, for a network comprised of several complete-graph-like SSGs, there will be several terms in the total spectrum entering in $\chi$. For $\underline\chi$, this yields a significantly lower value of the most highly degenerate eigenvalue, which will lead to a smaller value compared to the case for $\bm{A}[d(G)]$. 
For trees, the choice of the Hamiltonian was less crucial, since for $\bm{A}(G)$ as well as for $\bm{C}(G)$ the most highly degenerate eigenvalue is independent of the size of the star-like SSGs \cite{kulvelis2015universality}.

Since rings have a (at most) two-fold degenerate eigenvalues, we expect that in this case we obtain (nearly) optimal transport. We interpolate between these two limiting structures by a small-world-type mechanism \cite{watts1998collective}: By randomly removing a fixed fraction $1-p_{\text{int}}$ of internal bonds of each complete-graph-like SSG with size $n>3$ one obtains small-world-like SSGs (In order to be consistent with the usual small-world notation, we define $p_{\text{int}}$ with respect to the ring-like SSGs). Here, internal bonds are those bonds whose complete removal yields the ring-like SSG, see Fig.~\ref{fig.tree-dual}(b). For a complete graph of size $n$, there are $n(n-3)/2$ such internal bonds. 

We note, that the limiting structure of connected rings can lead to eigenvalues which are the same for different ring-like SSGs of the same size. Therefore, the collection of rings has a (slightly) larger value of $\chi$ compared to a single ring of the same size. Randomly adding bonds to the ring-like SSGs can easily result in lifting these accidental degeneracies, which will lead to a decrease of $\chi$. For a single ring, however, adding bonds always yields an increase in $\chi$ \cite{mulken2007quantum}.

\begin{figure}[b]
\centerline{\includegraphics[width=\columnwidth]{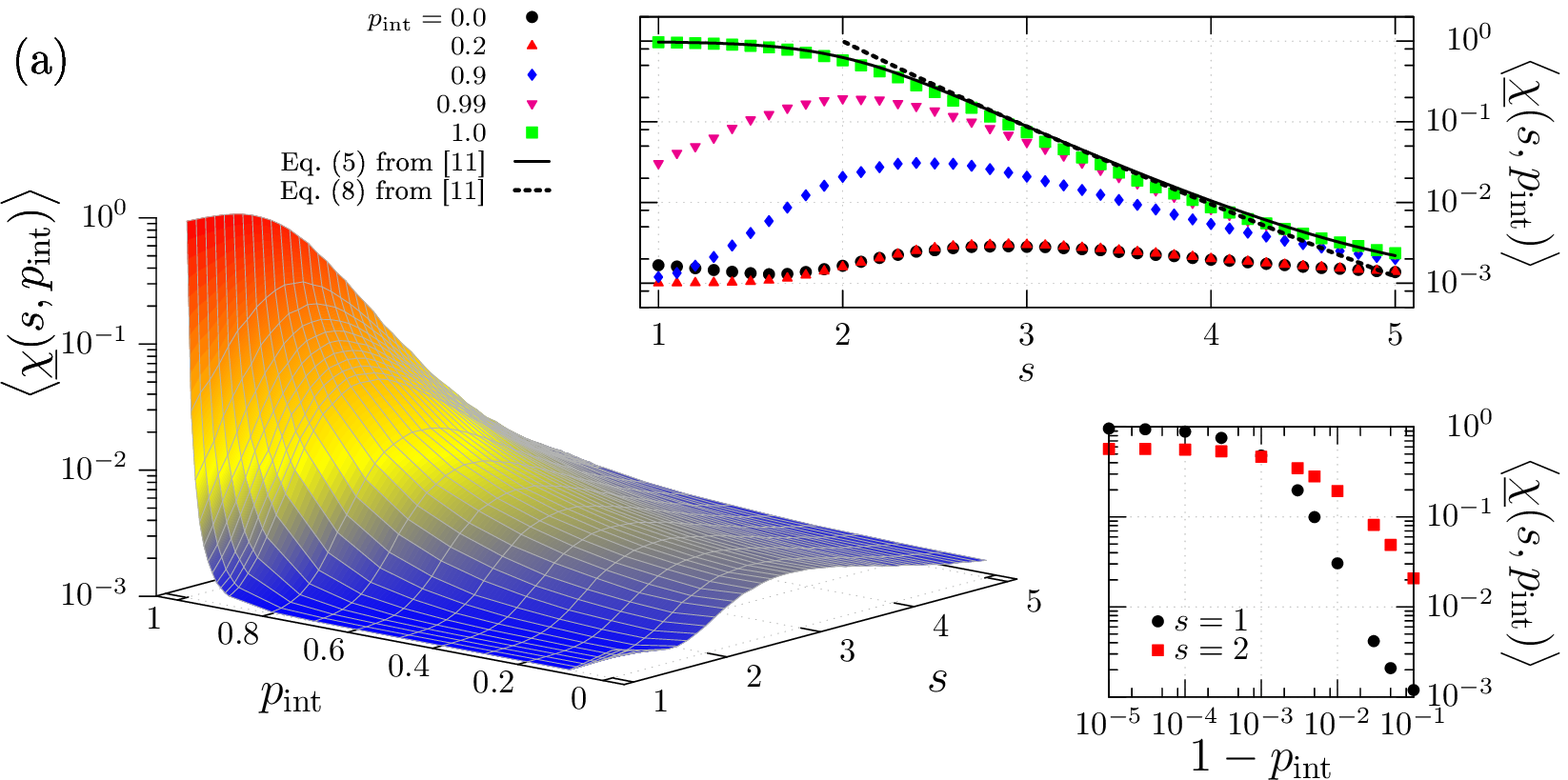}}
~\\
\centerline{\includegraphics[width=\columnwidth]{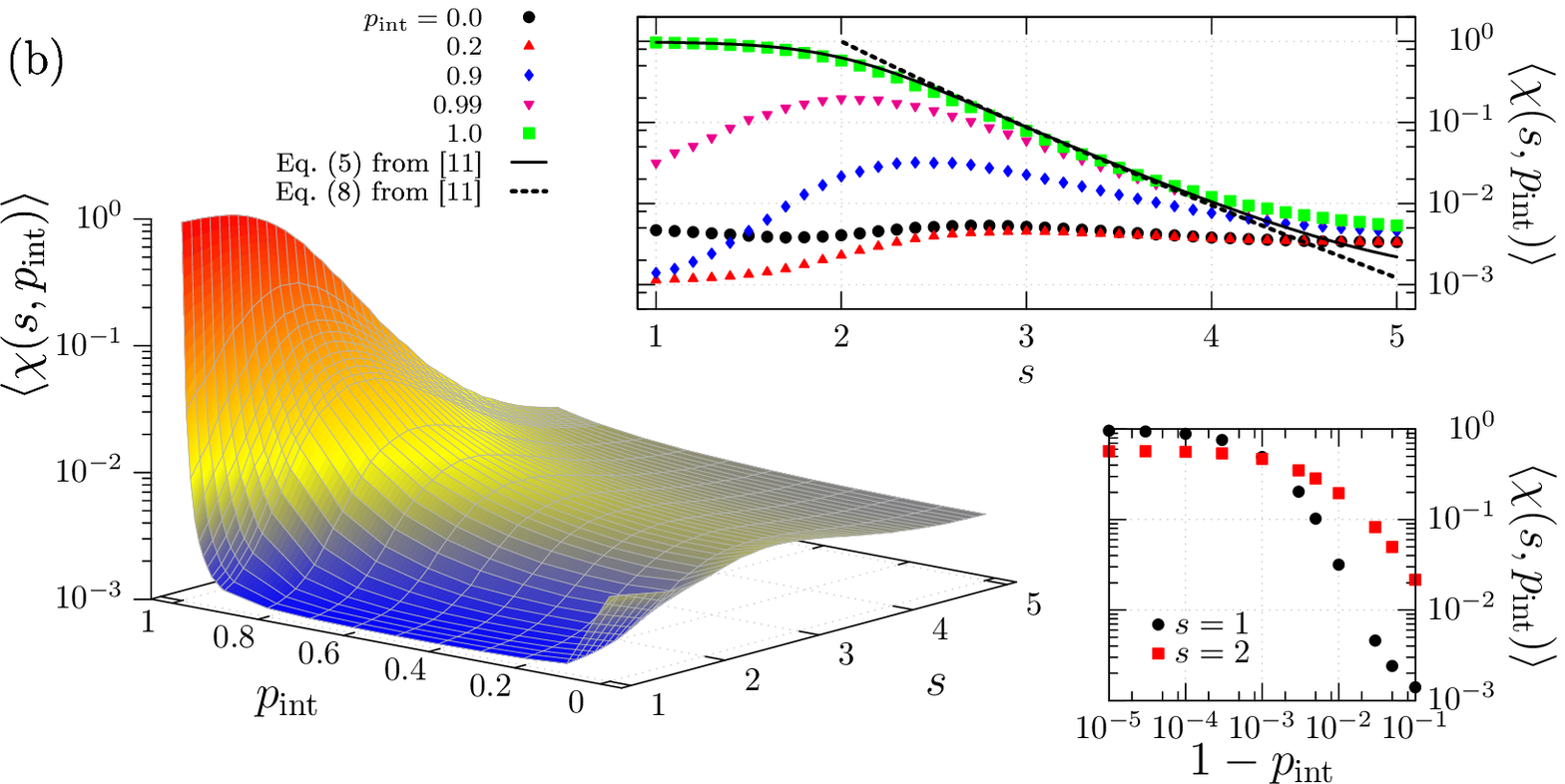}}
	\caption{(color online) Quantum transport properties of a complex network with loops and with SSG sizes chosen from a scale-free distribution. Numerical results (surface defined by dots) for the ensemble averaged global efficiency measures (a) $\avg{\underline\chi(s,p_{\text{int}})}$ and (b) $\avg{\chi(s,p_{\text{int}})}$ for $N=1000$ as function of the scale-free parameter $s$ and of the small-world-type parameter $p_{\text{int}}$ with $R=10^6/N$ realizations.  
Insets, upper panels: $\chi(s,p_{\text{int}})$ and $\underline\chi(s,p_{\text{int}})$ as a function of $s$ for five fixed values of $p_{\text{int}}$. The lines represent the analytic results given in Ref.~\cite{kulvelis2015universality} for $p_{\text{int}}=1$ and for $N=1000$ (solid) as well as for $N\to\infty$ (dashed) of the lower bound $\underline\chi$. The lower panels show $\chi(s,p_{\text{int}})$ and $\underline\chi(s,p_{\text{int}})$ for $s=1$ as a function of $1-p_{\text{int}}$ in double logarithmic scale.
}
\label{fig.phase_diagram}
\end{figure}

{\it Example.---}
We corroborate our findings and statements from above by numerical Monte-Carlo calculations of complex networks whose SSGs have a scale-free size distribution, $p_s(n) \propto n^{-s}$, with the sequential growth algorithm given in \cite{galiceanu2012relaxation}. 
Together with the small-world procedure, we have
$\chi=\chi(s,p_{\text{int}})$ and $\underline\chi=\underline\chi(s,p_{\text{int}})$ .

Figure~\ref{fig.phase_diagram} shows the transition from optimal transport to maximal breakdown of the ensemble averages of (a) $\avg{\underline\chi(s,p_{\text{int}})}$ and of (b) $\avg{\chi(s,p_{\text{int}})}$  as a function of the scaling parameter $s\in[1,5]$ and of the small-world-type parameter $p_{\text{int}}$ for networks of size $N=10^3$ with $R=10^6/N$ realizations. The insets show the $s$-dependence (upper right panels) for given values of $p_{\text{int}}$ and the $p_{\text{int}}$-dependence (lower right panel) for $s=1$ and $s=2$.
For $s\gtrsim5$, we find that there is only little difference between the two extreme cases of ring-like SSGs ($p_{\text{int}}=0$) and complete graph-like SSGs ($p_{\text{int}}=1$). In both cases, we obtain values of $\avg{\chi(5,0)}$ and of $\avg{\chi(5,1)}$ of the order of $1/N$. This is to be expected, since for large $s$ the dominant contribution to $p_s(n)$ comes from small values of $n$, leading to networks with long linear segments in both cases, resulting in rather flat $\varrho(E)$. Decreasing $s$ results in a separation of the two extreme cases: While the network with ring-like SSGs yields small values of $\avg{\chi(s,0)}$ for all $s\in[1,5]$, the network with complete-graph-like SSGs shows increasing values of $\avg{\chi(s,1)}$ indicating the onset of complete breakdown. Here, we also plot the analytic result of the lower bound $\avg{\underline\chi}$ for the corresponding tree (solid line in upper right panels) as well as the $N\to\infty$ result (dashed line in upper right panels), see Eqs.~(5) and (8) of Ref.~\cite{kulvelis2015universality}. As expected, the analytic result for the tree with $N=1001$ matches the data for the sequentially growing network for $p_{\text{int}}=1$ of size $N=1000$ (green squares). 

Starting from the complete-graph-like SSGs, $p_{\text{int}}=1$, we observe an initial decrease of $\avg{\chi(s,p_{\text{int}})}$ for all values of fixed $s$, which is related to smaller spectral densities of the highly degenerated eigenvalues. Especially, for small $s$ there is a sharp drop in $\avg{\chi(s,p_{\text{int}})}$ and $\avg{\underline\chi(s,p_{\text{int}})}$ for $p_{\text{int}}\lesssim1$. 
Small values of $s$, i.e., large $\avg{n}$ indicate a small number of SSGs, thus at least one SSG containing a significant fraction of all nodes. Assuming a single complete graph, one can estimate the change with $p_{\text{int}}$ in, at least, $\avg{\underline\chi(s,p_{\text{int}})}$: Removing a single bond from a complete graph will reduce the degeneracy of the highly degenerate eigenvalue by two because exactly two of the eigenvectors of the complete graph will not be eigenvectors of the modified graph. This change is small for large $N$, thus, indicating a smooth transition in $\avg{\underline\chi(s,p_{\text{int}})}$ as a function of $p_{\text{int}}$, see the lower right panels in Fig.~\ref{fig.phase_diagram}. A similar effect has been observed numerically for the transition from the complete graph to the star graph \cite{anishchenko2012enhancing}.
A single network with a small-world-type transition to the complete graph also allows for mean-field results of the spectrum \cite{grabow2012small}. There, one finds that the width of the spectrum becomes singular only for $p_{\text{int}}=1$, see Eq.~(8) of Ref.~\cite{grabow2012small}. For all other values of $p_{\text{int}}<1$, the eigenvalues are at most fwo-fold degenerate, indicating a discontinuous transition even for finite $N$. 

For small $s$, we also observe that already large values of $p_{\text{int}} <1 $ lead to values of $\avg{\underline\chi(s,p_{\text{int}})}$ and $\avg{\chi(1,p_{\text{int}})}$ which are below $\avg{\underline\chi(1,0)}$ and $\avg{\chi(1,0)}$, respectively. This supports our statement that incidental degeneracies are lifted by randomly adding bonds to the ring-like SSGs. Furthermore, only large values of $p_{\text{int}}$ yield highly degenerate eigenvalues of order $N$, see the differences for, say, $s=1$ between $p_{\text{int}}=0.9$ (diamonds in upper right panels) and $p_{\text{int}}=0.99$ (lower triangles).
Interestingly, for fixed $p_{\text{int}}\lesssim0.99$, maximal values of $\avg{\underline\chi(s,p_{\text{int}})}$ and $\avg{\chi(s,p_{\text{int}})}$ are found for $s\in[2,4]$, see also the upper panels in the insets in Fig.~\ref{fig.phase_diagram}. Thus, while the two limits of small and large $s$ lead to efficient transport, intermediate values of $s$ render the transport slightly less efficient. The maximal values of $\avg{\chi(s,p_{\text{int}})}$ and $\avg{\underline\chi(s,p_{\text{int}})}$ for fixed $p_{\text{int}}$ are, in fact, an artefact of the scale-free distribution $p_s(n)$: As is easily verified, this distribution leads to slightly increased probability to find $m$ SSGs of the same size $n$, $[p_s(n)]^m$, in the interval $s\in[2,4]$. 
This will cause accidental degeneracies because SSGs of the same size (can) lead to the same eigenvalues. 

{\it Conclusion.---}
We have shown that all sequentailly growing networks with single highly degenerate eigenvalues show universal behavior at the breakdown of quantum transport. The breakdown is driven by a parametric dependence of the spectral density of the network's Hamiltonian on the average size of the SSGs which themselves shown highly degenerate eigenvalues. 
Changing the topology of the SSGs from, say, complete-graph-like SSGs to ring-like SSGs allows to trigger 
a transition to optimal transport on complex quantum networks. Our general results are supported by numerical computations of complex quantum networks with a scale-free distribution of sizes of SSGs and a small-world-type transition to optimal transport.

\begin{acknowledgments}
We thank Alex Blumen, Niko Kulvelis, and Walter Strunz for fruitful discussion and valuable comments. Support from
the Deutsche Forschungsgemeinschaft is acknowledged by O. M. (DFG Grant No. MU2925/1-1) and by M. D. (DFG
Grants No. BL 142/11-1 and No. GRK 1642/1). M. G. acknowledges support from the Brazilian agency CNPq.
\end{acknowledgments}

\end{document}